\documentclass[doublecol]{epl2}
\begin{document}

\title{Quantitative features of multifractal subtleties in time series}

\author{Stanis{\l}aw Dro\.zd\.z\inst{1,2} \and Jaros{\l}aw Kwapie\'n\inst{1}
\and Pawe{\l} O\'swi\c ecimka\inst{1} \and Rafa{\l} Rak\inst{2}}
\shortauthor{S. Dro\.zd\.z \etal}

\institute{
\inst{1} Institute of Nuclear Physics, Polish Academy of Sciences, ul.
Radzikowskiego 152, PL - 31-342 Krak\'ow, Poland\\
\inst{2} Faculty of Mathematics and Natural Sciences, University of
Rzesz\'ow,
PL - 35-310 Rzesz\'ow, Poland }


\abstract{
Based on the Multifractal Detrended Fluctuation Analysis (MFDFA) and on the Wavelet Transform Modulus Maxima (WTMM) methods we investigate the origin of multifractality in the time series. Series fluctuating according to a $q$Gaussian distribution, both uncorrelated and correlated in time, are used. For the uncorrelated series at the border ($q=5/3$) between the Gaussian and the Levy basins of attraction asymptotically we find a phase-like transition between monofractal and bifractal characteristics. This indicates that these may solely be the specific nonlinear temporal correlations that organize the series into a genuine multifractal hierarchy. For analyzing various features of multifractality due to such correlations, we use the model series generated from the binomial cascade as well as empirical series. Then, within the temporal ranges of well developed power-law correlations we find a fast convergence in all multifractal measures. Besides of its practical significance this fact may reflect another manifestation of a conjectured $q$-generalized Central Limit Theorem.}

\maketitle

Multiscaling represents a concept that indicates an extremely promising direction towards grasping the essential characteristics of complexity. Indeed, the related measure in terms of multifractal spectra offers an attractively compact frame to quantify the hierarchy of scales and specificity of their interwoven organization. This in particular applies to the dynamical aspects of complexity. The empirics of such aspects is typically accessible through the time series containing information about the consecutive states of a complex system. A plenty of evidence has been reported that empirical data coming from such diverse fields as physics of turbulence~\cite{muzy91}, geophysics~\cite {dimitriu00}, astrophysics~\cite{abramenko05}, physics of plasma~\cite {burlaga92}, physiology~\cite{ivanov99}, complex networks research~\cite{bianconi01} and econophysics~\cite{fisher97} present signatures of multiscaling. As far as the time series are concerned an issue that however still remains open in the literature is which identifiable distinct attribute of complexity is in this context relevant. It seems commonly believed that either of the two, the nonlinear temporal correlations as well as abundantly accompanying them the non-Gaussian heavy tails of fluctuations, or both equally, may account for the multiscaling effects. Bearing in mind the very definition of the multifractal spectrum~\cite{halsey86} through its relation to the strength and organization of singularities in a series and a more recent work~\cite {muzy00} devoted to the construction and analysis of continuous multifractal cascades it seems however more natural to expect that these are the temporal correlations that are most relevant and even their scale free character demanded.

In order to quantitatively illuminate on the related issues we impose a selective discrimination by elaborating first of all the series with no time correlations at all, but instead with the systematically varying degree of heaviness of tails in their fluctuations. As an economic and physically motivated choice we use series sampled from the $q$Gaussian distribution
\begin{equation}
p (x) \sim e_q^{-a_qx^2}=1/[1+(q-1)a_qx^2]^{1/(q-1)} \label{qgaussians}
\end{equation}
whose origin can be routed to the nonextensive generalization of entropy~\cite{tsallis95,tsallis99}. The special $q=1$ case corresponds to the standard Gaussian distribution.

Further reason for considering this family of distributions is that they often well model the empirical data with heavy-tailed pdf's~\cite {upadhyaya01,rak07}. Moreover, according to the generalized Central Limit Theorem (CLT), the uncorrelated $q$Gaussian signals for $q > 1$ are unstable under convolution, and, depending on $q$, they are either in the standard Gaussian ($1 < q < 5/3$) or in the L\'evy stable distributions ($5/3 < q < 3$) basin of attraction with relation $\alpha_L = (3-q)/(q-1)$~\cite{tsallis95}. From this perspective an anticipated asymptotic for the uncorrelated series with $q$Gaussian pdf is a monofractal (Gaussian attractor) or a multifractal (L\'evy attractor)~\cite {jaffard99} which for the finite series assumes a bifractal form as for the truncated L\'evy processes~\cite {nakao00}.

At present the Multifractal Detrended Fluctuation Analysis (MFDFA)~\cite{kantelhardt02} and the Wavelet Transform Modulus Maxima (WTMM)~\cite{arneodo95} are the two most efficient practical methods to quantify multifractality. Both have their advantages, but the comparative evaluation of their overall performance~\cite{oswiecimka06} indicates that in most cases the more reliable choice is MFDFA. Therefore we use this method as our basic tool and occasionally WTMM (according to the procedure described for instance in ref.~\cite {oswiecimka06}) as an auxiliary independent consistency test.

The main points of MFDFA go as follows. Let ${x(i)}_{i=1,...,N}$ be a discrete signal. In MFDFA, one starts with the signal profile $Y(j) = \sum_{i=1}^j{(x(i)-<x>)}, \ j = 1,...,N$, where $<...>$ denotes averaging over all $i$'s. Then one divides the $Y(j)$ into $M_n$ non-overlapping segments of length $n$ ($n < N$) starting from both the beginning and the end of the signal ($2 M_n$ segments total). For each segment a local trend is estimated by fitting an $l$th order polynomial $P_{\nu}^{(l)}$, which is then subtracted from the signal profile. For the so-detrended signal a local variance $F^2(\nu,n)$ in each segment $\nu$ is calculated for each, from $n_{min}$ to $n_{max}$, scale variable $n$. Here, if not stated otherwise, we take $n_{min}=40$ and $n_{max}=N/5$. Finally, by averaging $F^2(\nu,n)$ over all segments $\nu$ one calculates the $r$th order fluctuation function:
\begin{equation}
F_r(n) = \bigg\{ \frac{1}{2 M_n} \sum_{\nu=1}^{2 M_n} [F^2(\nu,n)]^{r/2} \bigg\}^{1/r},
\label{ffunction}
\end{equation}
where $r \in \mathbf{R}$. Scaling behavior of the fluctuation function $F_r (n) \sim n^{h(r)}$ is an indication that the analyzed signal has fractal structure. If $h(r)= {\rm const}$, the signal is monofractal; it is multifractal otherwise. In both cases the singularity spectrum $f(\alpha)$~\cite{halsey86} can be calculated: $f (\alpha) =r [\alpha- h (r)] + 1$, where $\alpha=h(r)+r h'(r)$ is the singularity strength.

In the WTMM one calculates the wavelet transform $T_{\psi}(n,s)$ of a time series $x(i)$, where $\psi$ is the wavelet shifted by $n$, and $s$ is scale. For each scale one finds the local maxima of $T_{\psi}$ and then evaluates the partition function: $Z (r,s)=\sum_{l \in L(s)} {|T_{\psi}(n_l(s),s)|^{r}}$, where $L(s)$ denotes the set of all maxima for the scale $s$ and $n_l(s)$ stands for the position of a particular maximum. To preserve monotonicity of $Z(r,s)$ for different values of $s$, an additional criterion must be applied: $Z(r,s) = \sum_{l \in L(s)} {(\sup_{s'\le s} |T_{\psi}(n_l (s'),s')|)^ {r}}$. For fractal signals $Z(r,s) \sim s^{\ \tau(r)}$. The scaling exponent $\tau(r)$ is directly related with $\alpha$ and $f(\alpha)$ by the Legendre transform: $\alpha=\tau'(r)$ and $f(\alpha)=r \alpha - \tau(r)$.

Typically, the $f(\alpha)$ spectrum of a multifractal set resembles an inverted parabola which for monofractal signals reduces to a single point. In practice, however, one usually obtains a very narrow parabola even for the monofractal signals. The degree of multifractality can be evaluated by measuring width~\cite{kantelhardt02}
\begin{equation}
\Delta \alpha =\alpha_{\rm max}-\alpha_{\rm min} \label{deltaf}
\end{equation}
between the calculated extremal values of $\alpha$ for a given range $r_{\rm min} \le r \le r_{\rm max}$.

In order to assess the impact of a finite signal length on $f(\alpha)$ for the $q$Gaussian uncorrelated series, we generated 5 independent realizations of such series of different lengths: $N=10^4; 10^5; 10^6$, for a sequence of $q$ values from the range $1 \le q \le 2$ with a step $\Delta q=0.05$. This interval thus covers also the crossover regime near $q = 5/3$ between the Gaussian and the L\'evy attractor. The second- order polynomial as the detrending filter in MFDFA and the third derivative of Gaussian as the mother wavelet in WTMM are used. As we consider signals with heavy tails, the range of the index $r$ cannot be too broad; we choose $r \in [-4,4]$. This relatively narrow range of $r$ values allowed, automatically prevents a potential distortion of the results by the so-called "freezing" phenomenon which manifests itself in the linearization of $h(r)$ at large $r$ as observed for a log-normal cascade~\cite{lashermes04,muzy08}. This linearization may thus lead to an artificial narrowing of the singularity spectrum. In this connection it is also worth to notice - as a side remark - that as ref.~\cite{oswiecimka06} indicates the MFDFA procedure is less susceptible to such effects than WTMM.

$F_r(n)$ (MFDFA) calculated for a few exemplary choices of $q$ between 1.5 to 1.8 are shown in Figure~1 for $N=10^6$. For $q = 1.50$ (top left) there is a good power-law dependence of $F_r(n)$ for the whole range of scales and a small variability of the slope coefficients indicating a monofractal character of the analyzed data. On the opposite edge ($q \ge 1.70$, bottom row) one also observes a good scaling for all the scales and all the values of $r$, but here the slope coefficients are $r$-dependent. Between these two cases ($1.55 \le q \le 1.65$) there are signals with no uniform scaling of $F_r(n)$. Instead, we can distinguish two regions of different character: the small- scales region with the $r$-sensitive slope coefficients and the large-scales region without significant dependence on $r$. When going from $q=1.55$ to $q=1.70$, i.e. crossing the border between the Gaussian and the L\'evy regimes at $q=5/3$, the multiscaling region gradually extends towards the larger scales in the expense of the monofractal region. For shorter time series the crossover is more blurred. For the shortest considered signals it is at all difficult to distinguish the two regimes. One principal reason for these effects is that the scale parameter $n$ in Eq.~(\ref{ffunction}) needs to be sufficiently large so that through the requirements imposed on $F^2(\nu,n)$ by the CLT the fluctuation function sufficiently accurately probes the attractor.

%%%%%%%%%%%%%%% Figure 1 %%%%%%%%%%%%%%%%%
\begin{figure}[t]
\includegraphics[scale=0.35]{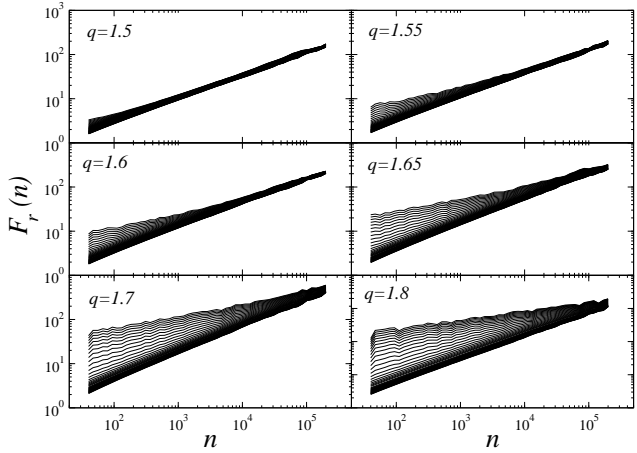}
\caption{Fluctuation function $F_r(n)$ (MFDFA) calculated for the signals of length $N=10^6$ sampled from different $q$Gaussians. In each panel $F_r(n)$ for different moments indexed by $r \in [-4,4]$ is displayed.}
\end{figure}
%%%%%%%%%%%%%%%%%%%%%%%%%%%%%%%%%%%%%

Figure 2 shows the width $\Delta\alpha(q)$ calculated with MFDFA (main panel) and with WTMM (inset). If two scaling regimes are observed in the $F_r(n)$ plots, the corresponding widths are calculated separately for each regime. Analytically calculated $f(\alpha)$ spectra for the finite signals in the L\'evy basin of attraction are bifractal and consist of only two points located at (0,0) and ($1/\alpha_L$,1) ~\cite {nakao00}. Numerically, such bifractal spectra cannot ideally be resolved and are represented by the shapes as in Figure~3. Thus the width in the main panel of Figure 2 is calculated from the continuous spectra for all the $q$-values and it is due to such effects that it assumes values slightly larger than the theoretical value $\Delta\alpha^{\rm (Levy)} = 1/\alpha_L$~\cite{oswiecimka06} for $q \ge 1.8$. If we approach the crossover at $q=5/3$, the deflection of $\Delta\alpha$ from $\Delta\alpha^{\rm (Levy)}$ becomes more significant. In the Gaussian basin of attraction all the considered signals doubtlessly appear monofractal only for $q \le 1.3 $, where the CLT convergence is fast. However, the signals with $N \ge 10^5$ are long enough for the CLT to exert its influence even for $q$ up to 1.5.

%%%%%%%%%%%%%%% Figure 2  %%%%%%%%%%%%%%%%%
\begin{figure}[t]
\includegraphics[scale=0.35]{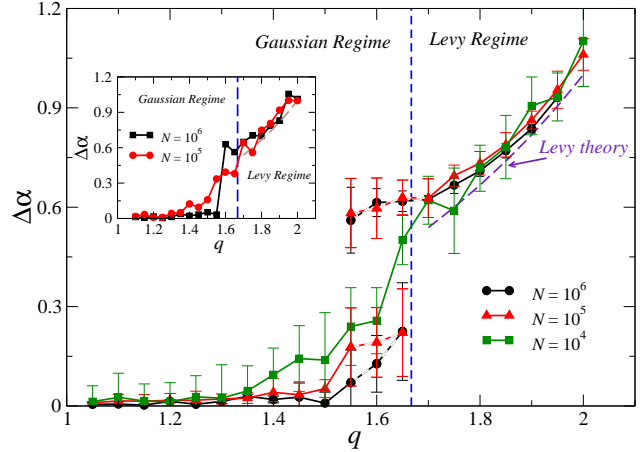}
\caption{The spectrum width $\Delta\alpha$ as a function of the $q$Gaussian parameter $q$. Dashed lines for $q < 5/3$ denote results obtained in the two competitive scaling regions of $F_r(n)$. The vertical dashed lines separate the Gaussian and the L\'evy basin of attraction. (Main panel) Results obtained with MFDFA. (Inset) Results obtained with WTMM. Error bars denote standard deviation of 5 independent realizations of the process.}
\label{delta_alfa}
\end{figure}
%%%%%%%%%%%%%%%%%%%%%%%%%%%%%%%%%%%%%

For $1.5 < q < 5/3$ the $q$Gaussians are significantly heavy-tailed, which makes their convergence to the Gaussian attractor very slow. This situation corresponds to the coexistence of two scaling regimes in Figure 1. The closer to the crossover point we are, the larger scale $n$ is needed to obtain the monofractal scaling in $F_r(n)$. In this region there exists no {\it a priori} criterion to choose one scaling regime and neglect the other, hence for such signals there is no unique value of $\Delta\alpha$. This problem may be resolved only by considering signals of much larger length (e.g., $N=10^9$), which can sometimes be generated for artificial signals, but are typically beyond reach for empirical data. The case of our shortest signals ($N=10^4$) illustrates how data with no actual scaling can erronously be considered suitable for calculating the scaling exponents. Inset of Figure 2 shows the same quantity but obtained from the wavelet-based analysis. Due to a restricted applicability of the WTMM method for shorter signals, only the $N=10^5$ and $N=10^6$ cases are displayed. These results are consistent with the ones by MFDFA.

%%%%%%%%%%%%%%% Figure 3 %%%%%%%%%%%%%%%%%%
\begin{figure}[t]
\includegraphics[scale=0.35]{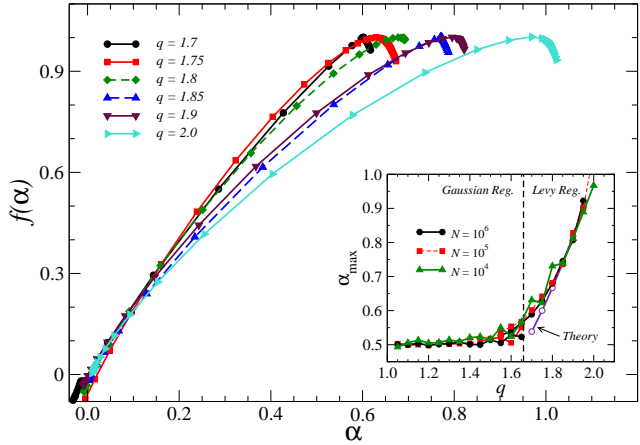}
\caption{(Main panel) MFDFA calculated singularity spectra $f(\alpha)$ for signals with $q$Gaussian pdf's for a few different choices of $q$ from the L\'evy basin of attraction. Analytical spectra, not shown here, are bifractal. (Inset) Inverse of the numerically calculated maximum $\alpha_{\rm max}$, which is directly related to the L\'evy parameter $\alpha_L$.}
\end{figure}
%%%%%%%%%%%%%%%%%%%%%%%%%%%%%%%%%%%%%%

In Figure 3 we show the $f(\alpha)$ spectra for signals from the L\'evy basin of attraction ($1.7 \le q \le 2.0$). They look multifractal but the high concentration of data points in two small regions near the points (0,0) and ($1/\alpha_L$,0) indicates the bifractal structure. Inset in Figure 3 displays $\alpha_{\rm max}(q)$. Outcomes of the numerical simulations for different $N$ almost follow the analytical prediction for $q \le 1.5$ (where $\alpha_{\rm max} \simeq 1/\alpha_L = 1/2$) and for $q \ge 1.8 $. In the vicinity of the threshold ($1.5 < q < 1.8$), the numerical value of $\alpha_ {\rm max}$ is larger than analytically expected.

Of an even greater interest are the series that involve correlations as well. As one example of such series, whose correlations are controllable and the asymptotic true result~\cite{calvet} is in addition known explicitly, we elect a signal generated by the log-normal cascade at the $k=20$ stage of cascading. This true result is typically well approximated already for $k \ge 12$. In the present case the cascade parameter $\lambda=1.1$ (the same notation as in ref.~\cite{oswiecimka06} is used). The temporal correlations in such a series decay uniformly on all the scales according to the power-law and, as using the cumulative $q$Gaussian distribution~\cite{rak07} is shown in Figure~4(c), the distribution of fluctuations at this stage ($k=20$) of cascading fits well to the $q$Gaussian with $q=1.6$. As before, the $N=10^4; 10^5; \sim 10^6$ (precisely $2^{20}$ in this last case) long series are extracted from the same $k=20$ cascade ($2^k$ data points), they thus obey the same distribution of fluctuations, $r \in [-4,4]$ and the scale variable $n \in [40,N/5]$. For $N=10^4$ and $N=10^5$ the result of calculations is averaged over the all nonoverlapping segments. The final result is shown in Figure~4(a). This result turns out essentially identical for different realisations of the above specified cascade. The convergence with $N$ of $f(\alpha)$ to the correct known result can be seen to be encouragingly fast. In the next step, destroying correlations in these series, by randomly shuffling their data points, results in fluctuation functions as in Figure~1 for $q=1.6$ with their characteristically increasing split in $r$ for the decreasing scale variables $n$. Consequently, extracting here the scaling exponents from $F_r(n)$ for $n$ varying within 40 up to say $2 \times 10^3$ interval - as by necessity would be done for a few thousand data points long series - results in $f(\alpha)$ shown in Figure~4(b). This outcome, largely even stable with increasing $N$, makes impression of revealing multifractality. By running the same procedure for $n$ from $2 \times 10^3$ up to the usual $N/5$ and increasing $N$ results in $f(\alpha)$ that systematically, though slowly, approaches an anticipated monofractal character.

%%%%%%%%%%%%%%% Figure 4 %%%%%%%%%%%%%%%%%%
\begin{figure}[t]
\includegraphics[scale=0.31]{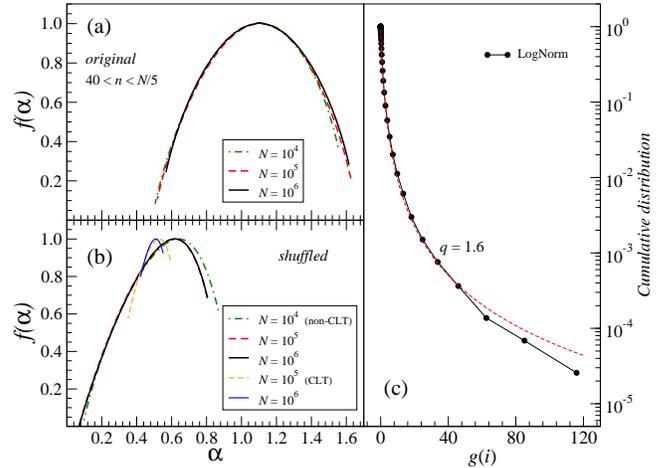}
\caption{(a) Singularity spectra $f(\alpha)$ for log-normal multiplicative cascades of different lengths. (b) $f(\alpha)$ for randomly shuffled cascade data, calculated for two different ranges of scaling variable: $40 < n < 2,000$ (``non-CLT regime'') and $20,000 < n < N/5$ (``CLT regime''). (c) Cumulative distribution function for the log- normal cascade (circles) with best-fitted $q$Gaussian distribution (dashed line).}
\end{figure}
%%%%%%%%%%%%%%%%%%%%%%%%%%%%%%%%%%%%%%

As a final example we elaborate the real empirical data. This is the time series representing the high frequency, 15-second logarithmic returns of the German stock market index (DAX) from the period November 28th, 1997 $-$ December 31th, 1999 with the trading hours 8:30 - 17:00 from the beginning which on September 20th 1999 until the end of this period was changed to 9:00 - 17:30. The first nonzero return for each day is removed from the analysis which results in $N=1,064,960$ data points long time series. The corresponding distribution of fluctuations fits also well by the $q$Gaussian with $q \approx 1.6$ which is shown in Figure~5(c). Exploring and quantifying the multiscaling characteristics in the financial time series is one of the mainstream issues in econophysics. Indeed, for our DAX series we obtain a satisfactory scaling of $F_r(n)$ which allows to evaluate $h(r)$. In order to permit making parallels with the cases considered above, we also divide the whole series into shorter pieces of length $N=10^5$ or $N=10^4 $ and average the results over all the pieces of the same length. The scaling variable $n \in [40,N/5]$ and the result is shown in Figure~5(a). At first glance somewhat unexpectedly (in view of the results for the log-normal cascade) one sees here no stable convergent result for $f (\alpha)$ in function of $N$. The span of the so-calculated $f(\alpha)$ systematically shrinks. Only a deeper inspection of the functional character of nonlinear correlations, as expressed for instance by the volatility (return modulus) autocorrelation, shown in Figure~6(a), opens perspective to understand this result. As expected, such an autocorrelation slowly decays according to the power-law. What however is not so expected in view of the present understanding of such effects thus far documented in the literature is that autocorrelation between events that are separated by more than about $2 \times 10^4$ basic units (here 15 seconds) suddenly drops down and starts oscillating between the positive and negative values with a decreasing amplitude (Figure~6 (b)). Such distant events are already not so consistently organized into a temporally power-law organized hierarchy relative to each other. Taking the scaling variable $n$ larger than the power-law correlation range thus introduces admixture of such a noise- like contribution which explains both the shrinkage of $f(\alpha)$ and a shift of its maximum towards $\alpha = 0.5$. That this is a relevant effect can be seen from Figure~5(b) where $f(\alpha)$ is calculated for $n$ up to $2 \times 10^4 $ independently on $N$. Now both $N=10^5$ and $N=10^6$ generate essentially the same result. We also verified that even going with $n$ up to only $2 \times 10^3$ already gives very similar result as the one that above is found as stable.

%%%%%%%%%%%%%%% Figure 5 %%%%%%%%%%%%%%%%%%
\begin{figure}[t]
\includegraphics[scale=0.31]{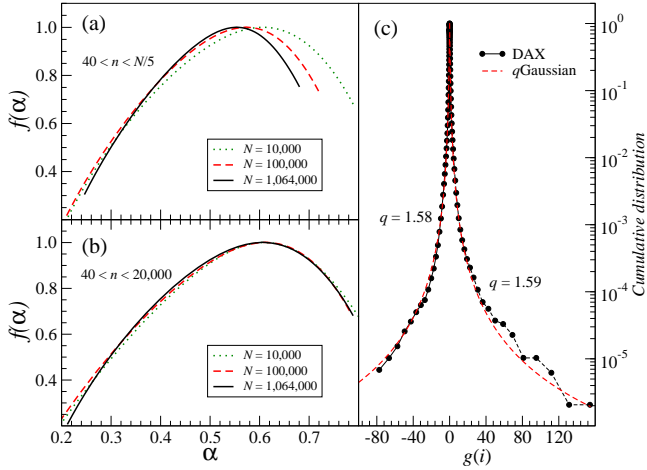}
\caption{Singularity spectra $f(\alpha)$ for empirical time series (German DAX 1 min returns) of different lengths, calculated in two different ranges of scale parameter: (a) $40 < n < N/5$ and (b) $40 < n < 20,000$. (c) Cumulative distribution function for DAX returns (circles) with $q$Gaussian distributions (dashed) best fitted separately for negative and positive returns.}
\end{figure}
%%%%%%%%%%%%%%%%%%%%%%%%%%%%%%%%%%%%%%

%%%%%%%%%%%%%%% Figure 6 %%%%%%%%%%%%%%%%%%
\begin{figure}[t]
\includegraphics[scale=0.31]{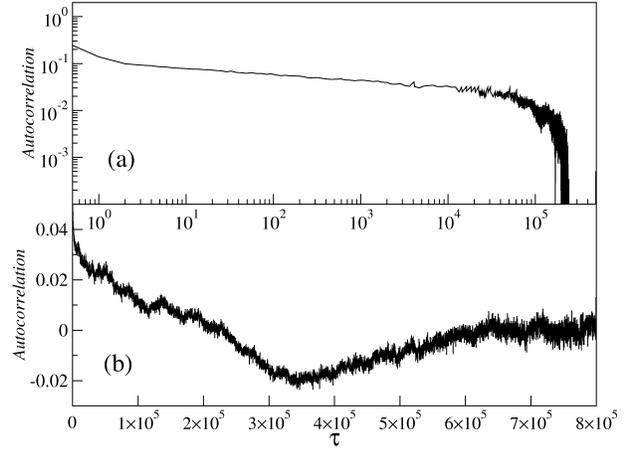}
\caption{Autocorrelation function for absolute DAX returns shown in log-log scale (a) and in the linear scale (b); note the region of negative autocorrelation values.}
\end{figure}
%%%%%%%%%%%%%%%%%%%%%%%%%%%%%%%%%%%%%%

The above results show that multifractality is a valuable but at the same time subtle concept. In the time series the genuine multifractality originates from temporal correlations. Encouragingly, within the range of well developed power-law temporal correlations, even in the most ``critical'' region of fluctuations just at the border between the Gaussian and the L\'evy basins of attraction, already a few thousand points long series suffices to properly evaluate $f(\alpha)$ using MFDFA. As an example with the financial data shows such an analysis needs however necessarily to be accompanied with evaluation of the range of those correlations and should be restricted to the intervals they consistently behave according to the power-law. In the realistic cases such correlations are likely to extend over the finite time-lag intervals only. Equivalently, a test of stability of the spectrum with the varying length of the series may serve the same purpose. Interestingly, much longer series are needed to reach convergence for the $q$Gaussian time-uncorrelated series, especially in the region characterized by $q$ close to $5/3$. In this latter case two results are asymptotically possible: either a monofractal singularity spectrum for $q < 5/3$ or a bifractal for $q > 5/3$ which means that no genuine multifractality is due to fluctuations. The observed faster convergence in the presence of power-law correlations may indicate another manifestation of the conjectured $q$-generalized CLT~\cite {tsallis06,hilhorst07}. Finally, the effects here elaborated and systematized, when taken into account, are expected to be crucially helpful in identifying the real multiscaling phenomena, their range and significance, and to reliably discern them from what often may turn out ``apparent''. Once a real multiscaling in this way is identified, in order to reach a higher level of confidence on the related quantitative characteristics, application of several by now available~\cite{palus08} statistical methods testing multifractality is recommended.

\end{document}